%% file: main.tex
\documentclass{article}

\usepackage{PRIMEarxiv}

\usepackage[utf8]{inputenc} 
\usepackage[T1]{fontenc}    
\usepackage{hyperref}       
\usepackage{url}            
\usepackage{booktabs}       
\usepackage{amsfonts}       
\usepackage{nicefrac}       
\usepackage{microtype}      
\usepackage{lipsum}
\usepackage{fancyhdr}       
\usepackage{graphicx}       

\usepackage{cite}
\usepackage{amsmath,amssymb,amsfonts}
\usepackage{graphicx}
\usepackage{textcomp}
\usepackage{xcolor}
\usepackage{listings}
\usepackage{hyperref}
\usepackage{url}
\usepackage{xspace}
\usepackage{enumitem}

\usepackage[version=3]{acro}
\input{J_acronyms}

\usepackage[normalem]{ulem}

\usepackage{cleveref}
\usepackage{algorithm}
\usepackage{algpseudocode}
\usepackage{comment}

\usepackage{todonotes}

\usepackage{pgfplots}

\pgfplotsset{width=7cm,compat=1.8}

\def\BibTeX{{\rm B\kern-.05em{\sc i\kern-.025em b}\kern-.08em
    T\kern-.1667em\lower.7ex\hbox{E}\kern-.125emX}}
    
\definecolor{blue}{rgb}{0,0,1.0}
\definecolor{darkgreen}{rgb}{0,0.44,0}
\definecolor{green}{rgb}{0,0.44,0}
\definecolor{darkred}{rgb}{0.44,0,0}
\definecolor{darkblue}{rgb}{0,0,0.64}
\definecolor{mygray}{rgb}{0.9,0.9,0.9}
\definecolor{mymauve}{rgb}{0.58,0,0.82}
\definecolor{myred}{rgb}{0.72,0.18,0.0} 
\definecolor{mygreen}{rgb}    {0.0,0.72,0.0} 
\definecolor{myblue}{rgb} {0.18,0.0,0.72} 
\definecolor{mycreme}{rgb}        {1.0,0.8,0.2} 
\definecolor{mygray}{rgb}{0.95,0.95,0.95}
\definecolor{darkgray}{rgb}{0.55,0.55,0.55}

\newcommand{\gemm}{{\sc gemm}\xspace}
\newcommand{\pe}{\mathrel{+\!\!=}}

\graphicspath{{media/}}     

\title{Toward Matrix Multiplication for Deep Learning Inference
       on the Xilinx Versal
}

\author{
  Jie Lei, Jos\'e Flich, Enrique S. Quintana-Ort\'{\i}\\
  Depto. de Informática de Sistemas y Computadores \\
  Universitat Politècnica de València \\
  Valencia, Spain\\
  \texttt{\{jlei,jflich,quintana\}@disca.upv.es} \\
}
\begin{document}
\maketitle
\begin{abstract}
The remarkable positive impact of Deep Neural Networks on many Artificial Intelligence (AI) tasks 
has led to the development of various high performance algorithms as well as specialized
processors and accelerators.
In this paper we address this scenario by demonstrating that the principles underlying
the modern realization of the general matrix multiplication (\gemm) in conventional processor
architectures, are also valid to achieve high performance 
for the type of operations that arise in deep learning (DL) on
an exotic accelerator such as the AI Engine (AIE) tile embedded in   
Xilinx Versal platforms. 
In particular, our experimental results with a prototype implementation of
the \gemm kernel, on a Xilinx Versal VCK190, delivers performance close to 
\textcolor{black}{86.7\%} of the theoretical peak that can be expected on an AIE tile, for \textcolor{black}{16-bit} integer
operands.
\end{abstract}


\input{sections/intro}
\input{sections/related}
\input{sections/xilinx}

\input{sections/gemm}
\input{sections/design}

\input{sections/results}

\pagebreak
 \section{Glossary}
\noindent
\textbf{ACAP}: Adaptive Compute Acceleration Platform;
\textbf{AIE}: Artificial Intelligence Engine;
\textbf{API}: Application Programming Interface;
\textbf{CNN}: Convolutional Neural Network;
\textbf{DDR}: Double Data Rate (memory);
\textbf{DL}: Deep Learning;
\textbf{DSP}: Digital Signal Processor;
\textbf{GMIO}: Global Memory Input/Output;
\textbf{INT8}: Integer 8-bit (arithmetic);
\textbf{LUT}: Look-Up Table;
\textbf{MAC}: Multiply-and-Accumulate;
\textbf{NoC}: Network-on-chip;
\textbf{PL}: Programmable Logic;
\textbf{SIMD}: Single-Instruction Multiple-Data;
\textbf{TOPS}: Tera-operations per second;
\textbf{TPU}: Tensor Processing Unit;
\textbf{\gemm}: General Matrix Multiplication;
\textbf{RAM}: Random Access Memory.

\section*{Acknowledgments}
The authors gratefully acknowledge funding from European Union’s Horizon2020 Research and Innovation programme under the Marie Skłodowska Curie Grant Agreement No. 956090 (APROPOS, http://www.apropos-itn.eu/).

This work also received funding in Spain from the 
research project
PID2020-113656RB-C22 of MCIN/AEI/10.13039/501100011033,
y por FEDER \textit{Una manera de hacer Europa}.
as well as from 
European High-Performance Computing Joint Undertaking (JU) under grant agreement No.
955558 (eFlows4HPC project). The JU receives support from the European Union’s Horizon 2020 research and innovation programme,
and Spain, Germany, France, Italy, Poland, Switzerland, Norway.

\bibliographystyle{IEEEtran}
\bibliography{J_BIB,E_BIB}

\end{document}

%% file: J_acronyms.tex
\DeclareAcronym{TPU}{
tag = {esr9},
short = {TPU},
long = {tensor processing unit},
pdfcomment = {tensor processing unit},
}

\DeclareAcronym{LUT}{
tag = {esr9},
short = {LUT},
long = {lookup table},
pdfcomment = {lookup table}
}

\DeclareAcronym{PL}{
tag = {esr9},
short = {PL},
long = {programmable logic},
pdfcomment = {programmable logic}
}

\DeclareAcronym{PS}{
tag = {esr9},
short = {PS},
long = {processing system},
pdfcomment = {processing system}
}

\DeclareAcronym{MB}{
tag = {esr9},
short = {MB},
long = {Mbyte},
pdfcomment = {mega byte}
}

\DeclareAcronym{KB}{
tag = {esr9},
short = {KB},
long = {Kbyte},
pdfcomment = {kilo byte}
}

\DeclareAcronym{GB}{
tag = {esr9},
short = {GB},
long = {Gbyte},
pdfcomment = {giga byte}
}

\DeclareAcronym{GEMM}{
tag = {esr9},
short = {GEMM},
long = {general matrix multiplication},
pdfcomment = {general matrix multiplication}
}

\DeclareAcronym{BLIS}{
tag = {esr9},
short = {BLIS},
long = {BLAS-like library instantiation software},
pdfcomment = {BLAS-like library instantiation software}
}

\DeclareAcronym{GMIO}{
tag = {esr9},
short = {GMIO},
long = {global memory input output},
pdfcomment = {global memory input output}
}

\DeclareAcronym{DDR}{
tag = {esr9},
short = {DDR},
long = {double data rate},
pdfcomment = {double data rate}
}

\DeclareAcronym{AI}{
    tag = {esr9},
    short = {AI},
    long = {artificial intelligence},
    pdfcomment = {artificial intelligence}
}

\DeclareAcronym{ANN}{
    tag = {esr9},
    short = {ANN},
    long = {artificial neural network},
    pdfcomment = {artificial neural network}
}

\DeclareAcronym{ASIC}{
    tag = {esr9},
    short = {ASIC},
    long = {application-specific integrated circuit},
    pdfcomment = {application-specific integrated circuit}
}

\DeclareAcronym{BD}{
    tag = {esr9},
    short = {BD},
    long = {bounded delay},
    pdfcomment = {bounded delay}
}

\DeclareAcronym{BC}{
    tag = {esr9},
    short = {BC},
    long = {breast cancer},
    pdfcomment = {breast cancer}
}

\DeclareAcronym{BNN}{
    tag = {esr9},
    short = {BNN},
    long = {binarized neural network},
    pdfcomment = {binarized neural network}
}

\DeclareAcronym{CB}{
    tag = {esr9},
    short = {CB},
    long = {connectionist bench},
    pdfcomment = {connectionist bench}
}

\DeclareAcronym{CD}{
    tag = {esr9},
    short = {CD},
    long = {completion detection},
    pdfcomment = {completion detection}
}

\DeclareAcronym{CNN}{
    tag = {esr9},
    short = {CNN},
    long = {convolutional neural network},
    pdfcomment = {convolutional neural network}
}

\DeclareAcronym{CTM}{
    tag = {esr9},
    short = {CTM},
    long = {convolutional Tsetlin machine},
    pdfcomment = {convolutional Tsetlin machine}
}

\DeclareAcronym{CPOG}{
    tag = {esr9},
    short = {CPOG},
    long = {conditional partial order graph},
    pdfcomment = {conditional partial order graph}
}

\DeclareAcronym{DI}{
    tag = {esr9},
    short = {DI},
    long = {delay insensitive},
    pdfcomment = {delay insensitive}
}

\DeclareAcronym{DR}{
    tag = {esr9},
    short = {DR},
    long = {dual-rail},
    pdfcomment = {dual-rail}
}

\DeclareAcronym{DRAM}{
    tag = {esr9},
    short = {DRAM},
    long = {dynamic RAM},
    pdfcomment = {dynamic RAM}
}

\DeclareAcronym{DSP}{
    tag = {esr9},
    short = {DSP},
    long = {digital signal processor},
    pdfcomment = {digital signal processor}
}

\DeclareAcronym{DVFS}{
    tag = {esr9},
    short = {DVFS},
    long = {dynamic voltage and frequency scaling},
    pdfcomment = {dynamic voltage and frequency scaling}
}

\DeclareAcronym{FD-SOI}{
    tag = {esr9},
    short = {FD-SOI},
    long = {fully-depleted silicon-on-insulator},
    pdfcomment = {fully-depleted silicon-on-insulator}
}

\DeclareAcronym{FPGA}{
    tag = {esr9},
    short = {FPGA},
    long = {field-programmable gate array},
    pdfcomment = {field-programmable gate array}
}

\DeclareAcronym{FSM}{
    tag = {esr9},
    short = {FSM},
    long = {finite state machine},
    pdfcomment = {finite state machine}
}

\DeclareAcronym{HDL}{
    tag = {esr9},
    short = {HDL},
    long = {hardware description language},
    pdfcomment = {hardware description language}
}

\DeclareAcronym{HVR}{
    tag = {esr9},
    short = {HVR},
    long = {house voting record},
    pdfcomment = {house voting record}
}

\DeclareAcronym{INWE}{
    tag = {esr9},
    short = {INWE},
    long = {inverse-narrow-width effect},
    pdfcomment = {inverse-narrow-width effect}
}

\DeclareAcronym{ISA}{
    tag = {esr9},
    short = {ISA},
    long = {instruction set architecture},
    pdfcomment = {instruction set architecture}
}

\DeclareAcronym{IoT}{
    tag = {esr9},
    short = {IoT},
    long = {Internet of Things},
    pdfcomment = {Internet of Things}
}

\DeclareAcronym{LA}{
    tag = {esr9},
    short = {LA},
    long = {learning automaton},
    pdfcomment = {learning automaton}
}

\DeclareAcronym{LEC}{
    tag = {esr9},
    short = {LEC},
    long = {logical equivalence checking},
    pdfcomment = {logical equivalence checking}
}

\DeclareAcronym{LFSR}{
    tag = {esr9},
    short = {LFSR},
    long = {linear feedback shift register},
    pdfcomment = {linear feedback shift register}
}

\DeclareAcronym{LU}{
    tag = {esr9},
    short = {LU},
    long = {learning unit},
    pdfcomment = {learning unit}
}

\DeclareAcronym{MAC}{
    tag = {esr9},
    short = {MAC},
    long = {multiply-and-accumulate},
    pdfcomment = {multiply-and-accumulate}
}

\DeclareAcronym{MNIST}{
    tag = {esr9},
    short = {MNIST},
    long = {Modified National Institute of Standards and Technology},
    pdfcomment = {Modified National Institute of Standards and Technology }
}

\DeclareAcronym{MEP}{
    tag = {esr9},
    short = {MEP},
    long = {minimum energy point},
    pdfcomment = {minimum energy point}
}

\DeclareAcronym{ML}{
    tag = {esr9},
    short = {ML},
    long = {machine learning},
    pdfcomment = {machine learning}
}

\DeclareAcronym{MLP}{
    tag = {esr9},
    short = {MLP},
    long = {multi-layer perceptron},
    pdfcomment = {multi-layer perceptron}
}

\DeclareAcronym{MPP}{
    tag = {esr9},
    short = {MPP},
    long = {maximum power point},
    pdfcomment = {maximum power point}
}

\DeclareAcronym{MPPT}{
    tag = {esr9},
    short = {MPPT},
    long = {maximum power point tracking},
    pdfcomment = {maximum power point tracking}
}

\DeclareAcronym{MRAM}{
    tag = {esr9},
    short = {MRAM},
    long = {},
    pdfcomment = {}
}

\DeclareAcronym{NCL}{
    tag = {esr9},
    short = {NCL},
    long = {null convention logic},
    pdfcomment = {null convention logic}
}

\DeclareAcronym{NN}{
    tag = {esr9},
    short = {NN},
    long = {neural network},
    pdfcomment = {neural network}
}

\DeclareAcronym{OCV}{
    tag = {esr9},
    short = {OCV},
    long = {on-chip variation},
    pdfcomment = {on-chip variation}
}

\DeclareAcronym{PVT}{
    tag = {esr9},
    short = {PVT},
    long = {process, variation and temperature},
    pdfcomment = {process, variation and temperature}
}

\DeclareAcronym{QDI}{
    tag = {esr9},
    short = {QDI},
    long = {quasi delay insensitive},
    pdfcomment = {quasi delay insensitive}
}

\DeclareAcronym{RAM}{
    tag = {esr9},
    short = {RAM},
    long = {random access memory},
    pdfcomment = {random access memory}
}

\DeclareAcronym{RDF}{
    tag = {esr9},
    short = {RDF},
    long = {random dopant fluctuation},
    pdfcomment = {random dopant fluctuation}
}

\DeclareAcronym{RTM}{
    tag = {esr9},
    short = {RTM},
    long = {recurrent Tsetlin machine},
    pdfcomment = {recurrent Tsetlin machine}
}

\DeclareAcronym{SCM}{
    tag = {esr9},
    short = {SCM},
    long = {standard cell memory},
    pdfcomment = {standard cell memory}
}

\DeclareAcronym{SI}{
    tag = {esr9},
    short = {SI},
    long = {speed independent},
    pdfcomment = {speed independent}
}

\DeclareAcronym{SR}{
    tag = {esr9},
    short = {SR},
    long = {single-rail},
    pdfcomment = {single-rail}
}

\DeclareAcronym{SRAM}{
    tag = {esr9},
    short = {SRAM},
    long = {static RAM},
    pdfcomment = {static RAM}
}

\DeclareAcronym{STG}{
    tag = {esr9},
    short = {STG},
    long = {signal transition graph},
    pdfcomment = {signal transition graph}
}

\DeclareAcronym{SVM}{
    tag = {esr9},
    short = {SVM},
    long = {support vector machine},
    pdfcomment = {support vector machine}
}

\DeclareAcronym{TA}{
    tag = {esr9},
    short = {TA},
    long = {Tsetlin automaton},
    pdfcomment = {Tsetlin automaton}
}

\DeclareAcronym{TAT}{
    tag = {esr9},
    short = {TAT},
    long = {Tsetlin automaton team},
    pdfcomment = {Tsetlin automaton team}
}

\DeclareAcronym{TM}{
    tag = {esr9},
    short = {TM},
    long = {Tsetlin machine},
    pdfcomment = {Tsetlin machine}
}

\DeclareAcronym{ULV}{
    tag = {esr9},
    short = {ULV},
    long = {ultra-low voltage},
    pdfcomment = {ultra-low voltage}
}

\DeclareAcronym{VLSI}{
    tag = {esr9},
    short = {VLSI},
    long = {very-large-scale integration},
    pdfcomment = {very-large-scale integration}
}

\DeclareAcronym{WSN}{
    tag = {esr9},
    short = {WSN},
    long = {wide sensor network},
    pdfcomment = {wide sensor network}
}

\DeclareAcronym{IC}{
    tag = {esr9},
    short = {IC},
    long = {integrated circuit},
    pdfcomment = {integrated circuit}
}

\DeclareAcronym{MFCC}{
    tag = {esr9},
    short = {MFCC},
    long = {Mel-frequency cepstrum coefficients},
    pdfcomment = {Mel-frequency cepstrum coefficients}
}

\DeclareAcronym{KWS}{
    tag = {esr9},
    short = {KWS},
    long = {keyword spotting},
    pdfcomment = {keyword spotting}
}

\DeclareAcronym{DNN}{
    tag = {esr9},
    short = {DNN},
    long = {Deep Neural Network},
    pdfcomment = {Deep Neural Network}
}

\DeclareAcronym{QCNN}{
    tag = {esr9},
    short = {QCNN},
    long = {Quantized Convolutional Neural Network},
    pdfcomment = {Quantized Convolutional Neural Network}
}

\DeclareAcronym{BCNN}{
    tag = {esr9},
    short = {BCNN},
    long = {Binarized Convolutional Neural Network},
    pdfcomment = {Binarized Convolutional Neural Network}
}

\DeclareAcronym{LTSM}{
    tag = {esr9},
    short = {LTSM},
    long = {Long tag = {esr9},
    short-term Memory},
    pdfcomment = {Long tag = {esr9},
    short-term Memory}
}

\DeclareAcronym{SoC}{
    tag = {esr9},
    short = {SoC},
    long = {system on a chip},
    pdfcomment = {system on a chip}
}

\DeclareAcronym{GPU}{
    tag = {esr9},
    short = {GPU},
    long = {Graphics Processing Unit},
    pdfcomment = {Graphics Processing Unit}
}

\DeclareAcronym{HPC}{
    tag = {esr9},
    short = {HPC},
    long = {High Performance Computing},
    pdfcomment = {High Performance Computing}
}

\DeclareAcronym{CUDA}{
    tag = {esr9},
    short = {CUDA},
    long = {Compute Unified Device Architecture},
    pdfcomment = {Compute Unified Device Architecture}
}

\DeclareAcronym{INT}{
    tag = {esr9},
    short = {INT},
    long = {integer},
    pdfcomment = {integer}
}

\DeclareAcronym{FP}{
    tag = {esr9},
    short = {FP},
    long = {floating point},
    pdfcomment = {floating point}
}

\DeclareAcronym{CPU}{
    tag = {esr9},
    short = {CPU},
    long = {central processing unit},
    pdfcomment = {central processing unit}
}

\DeclareAcronym{CMISS-NN}{
    tag = {esr9},
    short = {CMISS},
    long = {Common Microcontroller Software Interface Standard - Neurl Network},
    pdfcomment = {Common Microcontroller Software Interface Standard - Neurl Network}
}

\DeclareAcronym{SIMD}{
    tag = {esr9},
    short = {SIMD},
    long = {single-instruction multiple-data},
    pdfcomment = {single-instruction multiple-data}
}

\DeclareAcronym{NoC}{
    tag = {esr9},
    short = {NoC},
    long = {Network-on-chip},
    pdfcomment = {Network-on-chip}
}

\DeclareAcronym{DMA}{
    tag = {esr9},
    short = {DMA},
    long = {direct memory access},
    pdfcomment = {direct memory access}
}

\DeclareAcronym{RISC}{
    tag = {esr9},
    short = {RISC},
    long = {reduced instruction set computer},
    pdfcomment = {reduced instruction set computer}
}

\DeclareAcronym{AIE}{
    tag = {esr9},
    short = {AIE},
    long = {Artificial Intelligence Engine},
    pdfcomment = {AI Engine}
}

%% file: sections/intro.tex
\section{Introduction}

In the last two decades, the slow-down of Moore's law
and the end of Dennard scaling
has resulted in the adoption of multicore processors, 
followed by the raise of domain-specific 
accelerators
(e.g., NVIDIA tensor cores and Google TPUs) %
and asymmetric multicore designs (e.g., ARM big.LITTLE; Apple M1, M2; 
and Intel Alder Lake, Raptor 
Lake)~\cite{4785860,1050511,10.1145/3282307,10.1145/3140659.3080246},
Xilinx is not oblivious to the performance-power advantages of specialized hardware and
has responded by introducing the Versal AI Engine (AIE) Core, 
a design that comprises a set of compute engines, 
advanced I/O, and integrated DDR controllers, targeting an ample range of 
workloads~\cite{8875639,9220682}.
%

In response to this, we analyze how to map the general matrix multiplication (\gemm) on an
\textit{AIE-enabled} Xilinx Versal Adaptive Compute Accelerated Platform (ACAP). 
Our main motivation for targeting this computational kernel is that \gemm 
is the cornerstone upon which many
scientific and engineering applications are built.
Furthermore, in deep learning (DL)
training and inference with 
the popular CNNs (convolutional neural networks) %
and the recent transformers, %
most arithmetic can be cast in terms of \gemm~\cite{Bar22,Che06}.

In addressing the efficient realization of \gemm on the Xilinx Versal, we make
the following two major contributions:
\begin{itemize}
\item We demonstrate that the ideas underlying the modern realization of \gemm on conventional
      processor architectures, equipped 
      with a hierarchical multilayered memory and single-instruction
      multiple-data (SIMD) arithmetic units, carry over 
      to the AIE-enabled Xilix Versal ACAP. For this purpose, we map the matrix operands
      to the distinct levels of the memory hierarchy as well as develop an
      architecture-specific \textit{micro-kernel} for the AIE tile.
\item We customize our general design of the algorithm for the particular case of the
      Xilinx Versal VCK190. Furthermore, we perform a thorough experimental analysis on this platform, exposing the 
      performance caveats and the key role
      of the cache configuration parameters.
\end{itemize}
At this point, we emphasize that our design is presented for the Versal VCK190, 
but it is straight-forward to adapt them for other AIE-enabled Versal ACAPs.

The rest of the paper is structured as follows:
In Section~\ref{sec:hardware} we briefly review the Versal VCK190;
and in Section~\ref{sec:GEMM}, we revisit the modern implementation of \gemm.
Next, in Section~\ref{sec:mapping} we describe the design of a \gemm 
micro-kernel for the Versal AIE tile, and the mapping of the
\gemm algorithm into the memory organization of the AIE-enabled ACAP.
In Section~\ref{sec:results} we perform a complete experimental evaluation of the
proposed algorithm; and in Section~\ref{sec:remarks} we discuss the insights gained
from our study, and sketch a roadmap for ongoing and future work.
After that, we close the paper with a glossary of acronyms used during the text.

%% file: sections/related.tex
\section{Related Work}

Multiply-accumulate (MAC) operations account for roughly 80\% of the arithmetic in machine learning inference~\cite{CASTELLO2022102459}.
These MACs operations can often be aggregated into a \gemm kernel,
for example in the case of (CNNs)~\cite{Ramirez2022BLIS}, 
transformers~\cite{transformer_FPGA}, 
and long short-term memory (LSTM) networks~\cite{LSTM_FPGA_MATRIX}.
As a result, extensive research has focused on accelerating \gemm, 
by optimizing memory accesses and/or speeding up the kernel 
using hardware accelerators. 
BLIS \cite{BLIS_FIRST_ACM} provides significantly higher memory efficiency and has been combined with approximating computing \cite{ZeeMixDomMixPBLIS}, targeting CNN and working with SIMD based accelerators \cite{9756685,9235053,Ramirez2022BLIS}.
In this paper, we 
map the BLIS design into Xilinx Versal platform.

Matrix multiplication is a fundamental kernel for deep learning inference and training. For the particular case
of CNNs, the IM2COL transform~\cite{Che06} casts the expensive convolution operators in terms of a \gemm by
transforming the input activation tensor into a much larger matrix.
For many convolutional layers, the \gemm that results from the application of this transform presents one
large dimension and two small ones. From that point of view, it often offers substantial performance gains 
to develop specialized implementations
of \gemm to tackle these cases particular \cite{9235053}.

%% file: sections/xilinx.tex
\section{Overview of the Xilinx Versal VCK190 ACAP}
\label{sec:hardware}

\subsection{Heterogeneous architecture}

\begin{figure}[htbp]
    \centering
    \includegraphics[width=0.5\columnwidth]{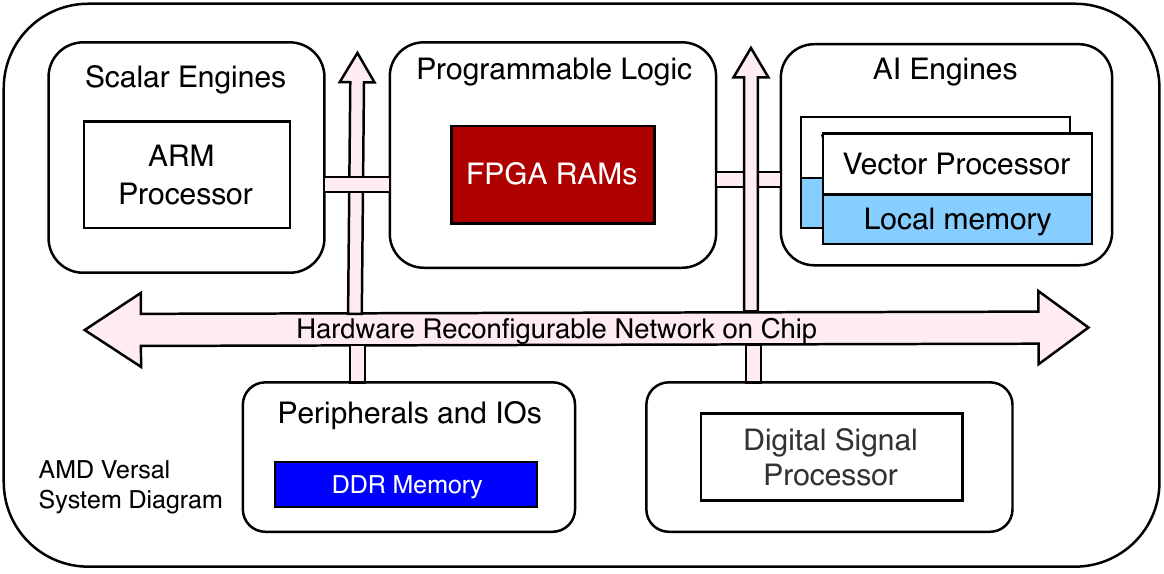}
    \vspace*{-2ex}
    \caption{Block diagram of the Versal AI Core.
    }
    \label{fig:acap}
\end{figure}

The 
 Versal AI Core comprises a collection of \textit{heterogeneous} silicon designs 
that aim to deliver high performance for both 
edge and cloud computing workloads; see
\Cref{fig:acap}. 
 For the particular case of Versal VC1902 processor~\cite{8875639} targeted in this work, the architecture contains:
\begin{enumerate}
\item A dual-core ARM Cortex-A72 processor
      for compute-intensive workloads,
      plus
      a dual-core ARM Cortex-R5F processor for 
      time-critical tasks.
\item 1,968 DSPs for accelerated arithmetic operations. 
 \item A customizable FPGA (or PL) 
       with 899,840 LUTs.
\item 400 standalone ``vector
      cores'', referred to as AIE tiles,
      and organized as a bidimensional array.
\end{enumerate}

Among other components, 
each AIE tile contains
a SIMD 
arithmetic unit and, altogether, these
vector cores conform  the VC1902's key feature for compute-intensive (DL) workloads. 
Concretely, they provide  up to 128 
MAC operations per clock cycle (per AIE
tile) for integer 8-bit (INT8) arithmetic~\cite{UG1079}.
Furthermore,    
the Versal VCK190 also contains a 
reprogrammable NoC
for high-bandwidth communication between the modules.

\subsection{Versatile memory architecture}
The Versal VCK190 
 features a flexible memory architecture basically consisting of
1) 32~KB of ``local'' memory per AIE tile; 
 2) distributed Block and Ultra RAMs 
with capacity for 4.3~MB 
and 16.3~MB, respectively; 
and 3) a global 2-GB DDR4 memory.
 Furthermore, each AIE tile  can directly access the 
32~KB of its ``neighbors'' in the 2D grid, while the
local memory of the remaining AIE tiles is reachable via
DMA.
The dedicated interconnects along with 
the NoC deliver non-blocking, 
deterministic communication for the AIEs.

The reconfigurability of the FPGA allows to leverage 
the Block RAM and Ultra RAM as temporary buffers  for the AIE, with the data being directly fed to the AIE tiles
via streaming interfaces for low latency access.  
 The AIE tiles access the global DDR4 via the NoC for higher throughput.

%% file: sections/gemm.tex
\begin{figure*}[t]
\centering
\begin{tabular}{ccc}
\begin{minipage}[t]{0.5\columnwidth}
\begin{tabular}{c}
\includegraphics[width=0.8\columnwidth]{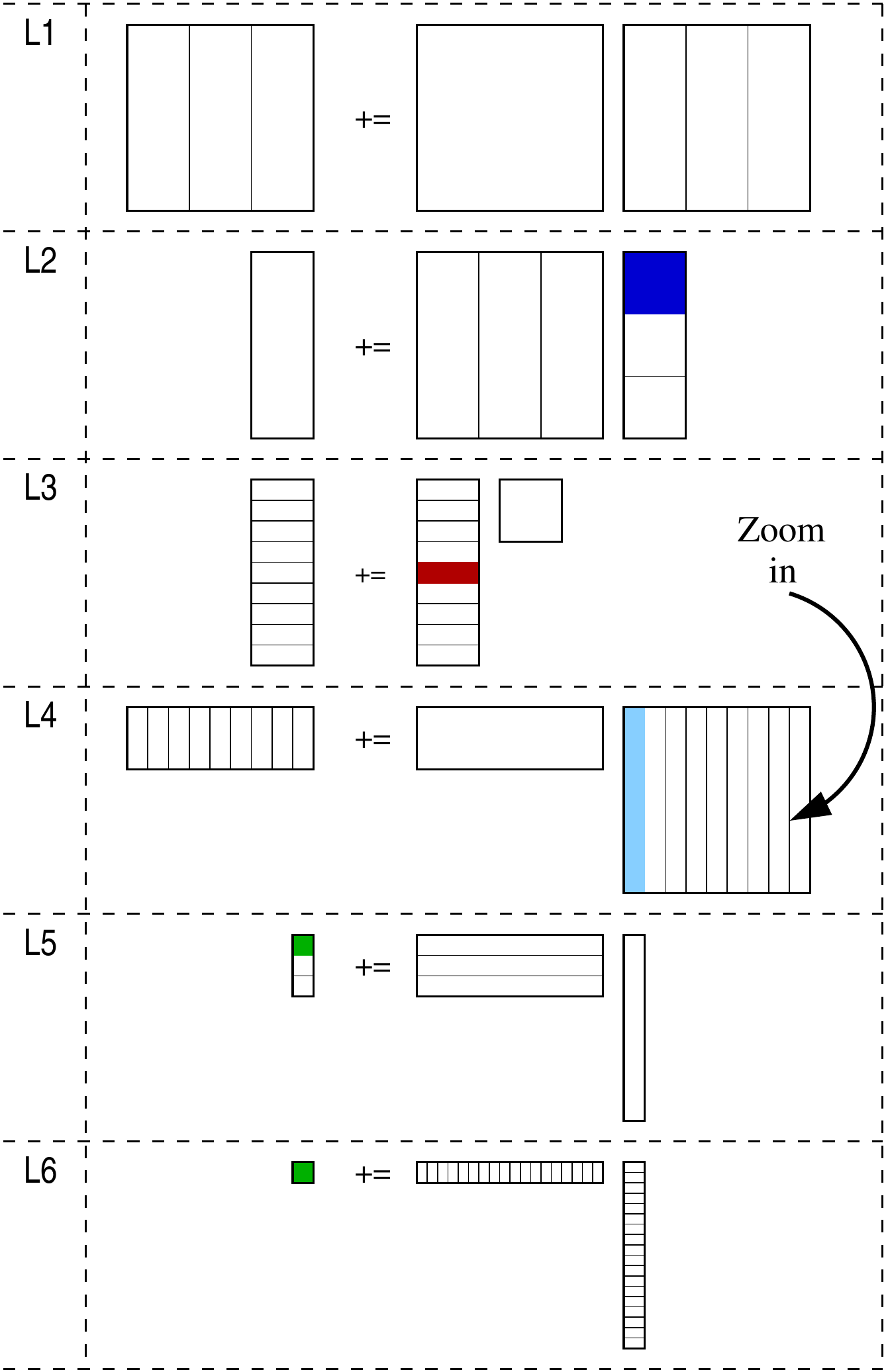} \\
~\\
\includegraphics[width=0.5\columnwidth]{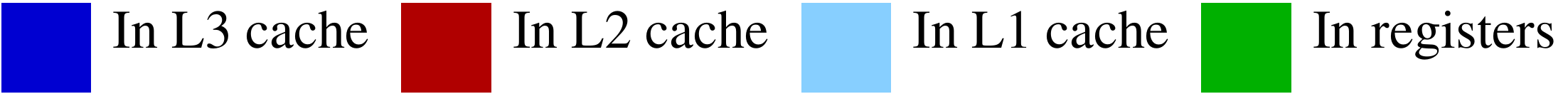}
\end{tabular}
\end{minipage}
&~~&
\begin{minipage}[t]{0.5\columnwidth}
\begin{tabular}{l}
\footnotesize
\begin{lstlisting}[language=C,alsoletter={.},deletekeywords={.sum},morekeywords={}]
for (jc=0; jc<n; jc+=nc)             // Loop L1
 for (pc=0; pc<k; pc+=kc) {          // Loop L2
  // Pack B
  Bc := B(pc:pc+kc-1,jc:jc+nc-1);
  for (ic=0; ic<m; ic+=mc) {         // Loop L3
   // Pack A
   Ac := A(ic:ic+mc-1,pc:pc+kc-1);
   for (jr=0; jr<nc; jr+=nr)         // Loop L4
     for (ir=0; ir<mc; ir+=mr)       // Loop L5
      // Micro-kernel
      C(ic+ir:ic+ir+mr-1, 
        jc+jr:jc+jr+nr-1)
        += Ac(ir:ir+mr-1,0:kc-1)
        *  Bc(0:kc-1,jr:jr+nr-1);
  }}
\end{lstlisting}
~\\
~\\
~\\
\footnotesize
\begin{lstlisting}[language=C,alsoletter={.},deletekeywords={.sum},morekeywords={}]
for (pr=0; pr<kc; pr++)              // Loop L6
  C(ic+ir:ic+ir+mr-1,
    jc+jr:jc+jr+nr-1)
    += Ac(ir:ir+mr-1,pr)
    *  Bc(pr,jr:jr+nr-1);
\end{lstlisting}
~\\
~\\
~\\
\includegraphics[width=0.7\columnwidth]{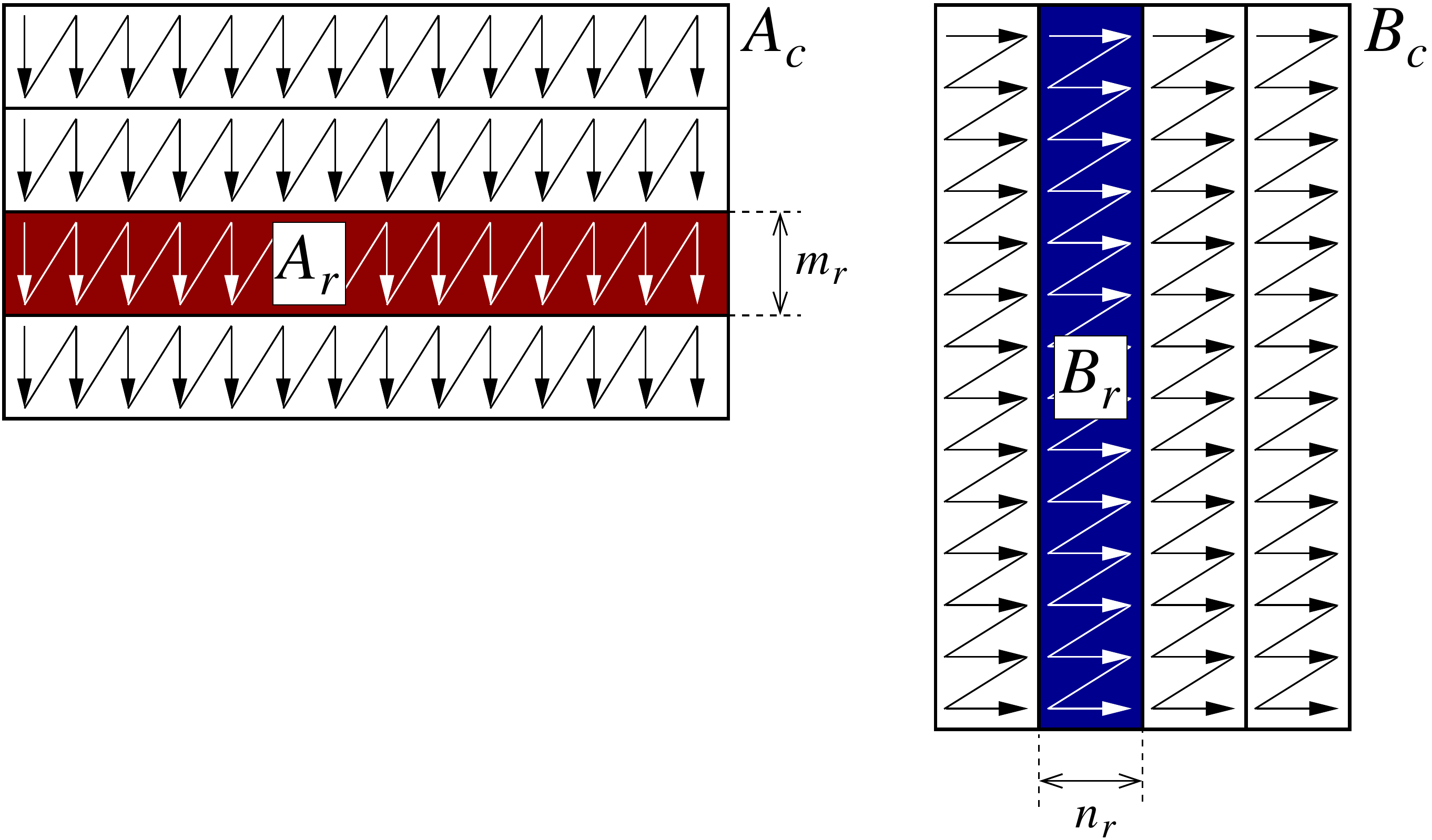}
\end{tabular}
\end{minipage}
\end{tabular}
\caption{Baseline high performance algorithm for \gemm. Left: data transfers across the memory hierarchy; Top-Right: blocked algorithm; Middle-Right: Micro-kernel; Bottom-Right: Packing of input matrix operands.}
\label{fig:baseline_GEMM}
\end{figure*}

\section{High Performance \gemm in connection with Deep Learning Inference}
\label{sec:GEMM}

\subsection{\gemm in conventional processor architectures}

Consider the \gemm
$C \pe AB$, where 
matrix $A$ is $m \times k$,
matrix $B$ is $k \times n$, and %
matrix $C$ is $m \times n$.
Modern implementations of this kernel (e.g., those
in AMD AOCL, OpenBLAS, BLIS and, 
possibly, Intel oneAPI) follow the ideas underlying 
GotoBLAS2~\cite{Goto:2008:AHP} to
apply blocking (tiling)~\cite{Dowd98} 
to the matrix operands using five nested
loops 
around two packing routines and an architecture-dependent %
\textit{micro-kernel}; see~\Cref{fig:baseline_GEMM} top-right,
and the loops labeled as \textsf{L1},\textsf{L2},\ldots,\textsf{L5} there.

For this baseline algorithm, a proper selection of the strides 
(also known as the \textit{cache configuration parameters})
for the three outermost loops,
given by $m_c,n_c,k_c$,
combined with a careful arrangement of the matrix inputs into
two buffers, $A_c,B_c$ (respectively, of dimensions
$m_c \times k_c$ and $k_c \times n_c$;
see Figure~\ref{fig:baseline_GEMM}, left and top-right),
orchestrated via the packing routines,
substantially reduces the number
of number of cache misses~\cite{BLIS1,BLIS4}.
For simplicity, hereafter we assume that 
$m,n,k$ are integer multiples of $m_c,n_c,k_c$, respectively.

In addition, the micro-kernel comprises one more loop
(labeled as \textsf{L6} in \Cref{fig:baseline_GEMM}, middle-right) 
that repeatedly updates an $m_r \times n_r$ 
\textit{micro-tile} of
$C$, known as $C_r$, via a sequence of $k_c$ rank-1 transforms, each involving 
a column of an $m_r \times k_c$ micro-panel of 
$A_c$ and a row of a $k_c \times n_c$ micro-panel of $B_c$. 
These micro-panels correspond to the blocks of $A_c,B_c$ denoted as 
$A_r,B_r$ in \Cref{fig:baseline_GEMM}, bottom-right.
In current processors with SIMD arithmetic units, the micro-kernel ``dimensions'', 
given by $m_r \times n_r$, are chosen to accommodate vectorization inside
the micro-kernel loop.
In this line, the specialized arrangement introduced by the packing 
routines
ensures accessing the data in the micro-panels
with unit stride from the micro-kernel which, in turn,
enables loading their data using SIMD instructions.

In summary, for high performance this formulation of \gemm relies on three parameters
($m_c,n_c,k_c$) which can be adjusted to fit the hierarchical memory
system of current architectures (see Figure~\ref{fig:baseline_GEMM} left), 
plus an architecture-dependent micro-kernel
that can be tuned, for example, to accommodate vectorization. 

\subsection{\gemm and DL inference}

A significant part of the arithmetic costs of recent transformers for natural language
processing is due to \gemm of the form 
$C \pe AB$, where $A$ is the weight matrix and $B/C$ are the input/output activation matrices.
In addition, for the popular CNNs leveraged in computer vision and general signal processing tasks, 
the lowering approach~\cite{Che06} casts
the convolution operator in terms of a large \gemm where
$A$ is the filter matrix, $C$ corresponds to the output activation matrix and
$B$ is the augmented matrix
obtained by applying the IM2COL (or IM2ROW) transform to the input activation matrix.
To tackle its large memory costs, 
this transformation can be blocked, as part of the packing done inside
the realization of the \gemm~\cite{9235053}, or applied on-the-fly, as part of the direct
convolution algorithm~\cite{Bar22}.

%% file: sections/design.tex
\section{Mapping \gemm to Xilinx Versal VCK190 for Deep Learning Inference}
\label{sec:mapping}

\subsection{Distributing the data across the memory hierarchy}

The Versal VCK190 offers a variety of possibilities for distributing the problem data 
across its memory resources
in order to attain high data reusability.
In particular, since the distinct memory levels in the Versal VCK190 
present different bandwidth rates and capacities, 
(see~\Cref{tab:multi_level_cache},)
depending on the arithmetic intensity of the target algorithm, 
this feature can be exploited to improve efficiency via
cache-friendly programming techniques, such as blocked algorithms~\cite{Dowd98},
combined with a careful distribution of the problem data across the levels of the memory hierarchy.

For the Versal VCK190, 
on the one hand,
the DDR4 memory provides sufficient space to accommodate
the global program data which, in our problem, corresponds to the \gemm matrix operands
$B,C$ comprising the input/output activations in the case of DL inference.
On the other hand,
the RAMs in the programmable logic provide high memory throughput, and can be employed as 
a scratchpad to temporarily store the full pre-packed matrix~$A$
for the weight/filter matrix. 
In practice, this implies that there is no need to keep a copy of the unpacked $A$ in global memory.
Finally, the fast and near-processor local memory delivers 
high throughout, and can be used to keep a packed micro-panel $B_r$.
These three levels, global memory, FPGA RAMs and local memory,
will play the role of the L1, L2 and L3 cache memories in
a conventional processor (see Figure~\ref{fig:baseline_GEMM} left),
as discussed next. %
\begin{table}[htb]
    \centering
    \caption{Multi-level memory hierarchy in the Versal VCK190.}
    \begin{tabular}{|l||r|}
    \hline
        Level  & Capacity %
        \\
    \hline \hline
        AIE vector registers    &  2~KB  \\ 
        AIE tile local memory   &  32~KB \\
        FPGA RAMs               &  20~MB \\
        DDR4 (global) memory    &  ~2~GB  
    \\\hline      
    \end{tabular}
    \label{tab:multi_level_cache}
\end{table}

\begin{figure*}[tbp!]
    \centering
    \includegraphics[width=0.8\textwidth]{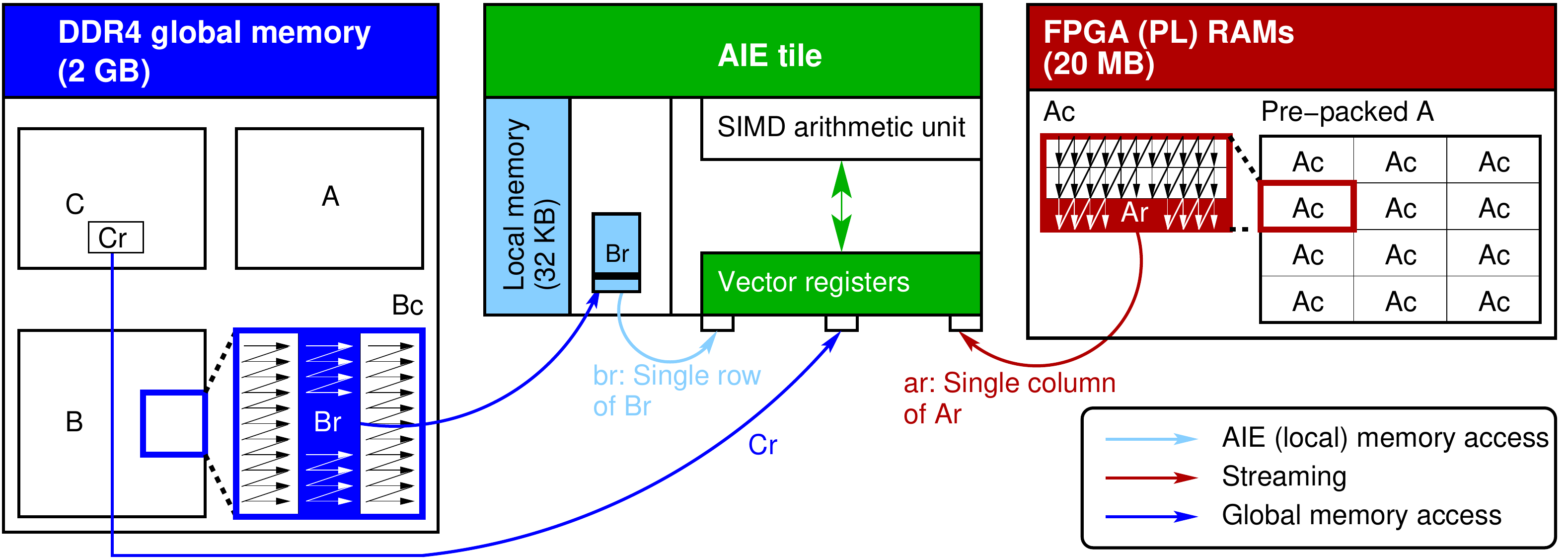}
    \caption{Data mapping and transfers between different memories of Versal VCK190.
    The data transfers from the micro-tile $C_r$ in the global memory, 
    the micro-panel $A_r$ in the FPGA memory, 
    and the micro-panel $B_r$ in the local memory move data directly to
    the vector registers. The data copy of a micro-panel of $B_C$ in the 
    global memory to $B_r$ in the local memory is carried out by one of the scalar
    engines (ARM processors).}
    \label{fig:systemDiag}
\end{figure*}

In \Cref{fig:systemDiag}, we graphically 
illustrate the distribution of the matrix operands and 
packed buffers adopted for mapping the \gemm kernel on the Versal VCK190. 
For simplicity, hereafter we consider a CNN model, and the \gemm that results from applying
the IM2ROW transform to a single convolution operator,
taking into account the special usage of \gemm
in a DL inference scenario to reduce the communication overhead:
\begin{itemize}
\item  The operand $A$ contains the read-only filters for the convolution.  
       The full matrix 
       can be thus pre-packed (off-line) 
       into a collection of buffers $A_c$ and kept 
       in the FPGA RAMs. Here we exploit that, 
       once a DL model is deployed, it is used to perform inference with many 
       input samples. Furthermore, for CNN models, the FPGA RAMs 
       are usually large enough to contain
       the filter parameters (weights and biases) for all the model. Therefore, 
       we can pre-pack $A$ and pre-load the result into the FPGA storage, 
       with a null cost for this during the inference process.
\item Matrix $B$ corresponds to the activation inputs of the convolution operator.
      This operand varies from one 
      inference sample to another and, therefore, 
      we need to pack it into buffer $B_c$ during
      the execution of \gemm. Following the approach in high performance realizations of this operation,
      we perform this packing in the DDR4 global memory (as there is no equivalent
      of an L3 cache in the Versal ACAP). 
      Also during the execution, the individual micro-panels $B_r$, 
      are copied into the AIE local memory via GMIO. For this purpose, the AIE tile directly accesses the global memory through the NoC.
      Here we amortize the cost of transferring $B_r$ by re-using the entries of this block 
      multiple times; %
      see loop \textsf{L5} in Figure~\ref{fig:baseline_GEMM}; top-right.
\item Finally, matrix  $C$ contains the activation outputs of the convolution operator. At each execution of the micro-kernel,
      a small micro-tile $C_r$, of dimension $m_r \times n_r$, 
      is first loaded directly from the DDR4 global memory into the 
      AIE tile vector registers, 
      and written back at the end of  the micro-kernel; 
      see Figure~\ref{fig:baseline_GEMM}, left and middle-right.
\end{itemize}

At this point, it is worth remarking some differences between 
our solution and the approach taken when implementing \gemm in a conventional processor:
\begin{itemize}
\item From the point of view of hardware, a conventional processor integrates a number of cache levels (usually, between two and three), and relies on a memory controller to orchestrate the data movements across them. In contrast, in the Versal VCK190 the transfers need to be explicitly encoded into the algorithm. This puts an extra burden into
the programmer's shoulders, but offers a strict control over the data transfers as the intermediate memory levels behave
as scratchpads. %
\item The Versal system provides (window and) streaming access methods for communication with
      the ``outside world.''
      Streaming provides high throughput and lower latency for the PL to AIE communication,
      but is also fundamentally different from the type of communication that occurs between
      the arithmetic units and the cache/memory levels of a conventional processor.
\item From the algorithmic point of view, given that the target is to implement \gemm for DL inference, we can pre-pack
      the weight/filter matrix $A$ into a collection of buffers residing into the FPGA RAM. Therefore, there is no need for
      the packing that occurs within loop \textsf{L3} of the 
      baseline algorithm for \gemm. %
\end{itemize}

\subsection{Design of the micro-kernel}
\label{subsec:microkernel}

Each AIE tile has four accumulator registers that can be 
used to store the results of the vector data path. 
An accumulator register is 768-bit wide and can be viewed as 16
accumulator lanes of (768/16=) 48 bits each. 
For high performance, the width of the accumulator registers affects
the practical dimensions of the
micro-kernel.
For the implementation of a DL-oriented \gemm in the Versal VC1902, we adopt
\textcolor{black}{INT16} as the baseline datatype (to be discussed later, in 
subsection~\ref{subsec:datatype}), and set the dimensions of the micro-kernel to 
$m_r \times n_r = 16 \times 4$; that is, the micro-tile $C_r$ 
updated by loop \textsf{L6} of the micro-kernel comprises 
16 entries (capacity of an accumulator register) for
4~columns (number of accumulator registers) of $C$.

\begin{figure}[tbh!]
\begin{flushright}
\begin{minipage}[t]{1\columnwidth}
\begin{lstlisting}[language=C++,alsoletter={.},deletekeywords={.sum},morekeywords={v32int16,v16int16,v16acc48,input_window_int16,int16,output_window_int16},numbersep=6pt]
#define Bref(i,j) Br[i*nr+j]
void micro_kernel( input_window_int16 * __restrict DDR_IN,
                   input_window_int16 * __restrict PL_IN,
                   int16 *Br,
                   output_window_int16 *__restrict out) {

  // Vectors, Accumulator for ar, br, C
  v32int16 ar0, ar1;
  v16int16 br;
  v16acc48 Cacc0, Cacc1, Cacc2, Cacc3;

  // Parameters for mac16() intrinsics
  unsigned int xoffsets    = 0x73727170;
  unsigned int xoffsets_hi = 0x77767574;
  unsigned int xsquare     = 0x3120;
  unsigned int zoffsets    = 0x0;
  unsigned int zoffsets_hi = 0x0;

  for (unsigned int i=0; i<Kc/4; i+=1) 
    // Unroll loop to overlap transfers and computations
    chess_prepare_for_pipelining    
    chess_loop_range(Kc/4,)  {
    // Read a vector of 16 INT16 into br
    br = *(v16int16*) Bref[i];

    // Read into ar
    ar0 = upd_w( ar0 ,0, window_readincr_v16(PL_IN));
    ar0 = upd_w( ar0 ,1, window_readincr_v16(PL_IN));
    ar1 = upd_w( ar1 ,0, window_readincr_v16(PL_IN));
    ar1 = upd_w( ar1 ,1, window_readincr_v16(PL_IN));

    // Compute two mac16()
    Cacc0 = mac16(Cacc0, ar0 ,0, xoffsets, xoffsets_hi,  xsquare, br ,0 , zoffsets, zoffsets_hi,4);
    Cacc1 = mac16(Cacc1, ar0 ,0, xoffsets, xoffsets_hi,  xsquare, br ,1 , zoffsets, zoffsets_hi,4);
    Cacc2 = mac16(Cacc2, ar0 ,0, xoffsets, xoffsets_hi,  xsquare, br ,2 , zoffsets, zoffsets_hi,4);
    Cacc3 = mac16(Cacc3, ar0 ,0, xoffsets, xoffsets_hi,  xsquare, br ,3 , zoffsets, zoffsets_hi,4);
    // 
    Cacc0 = mac16(Cacc0, ar1 ,0, xoffsets, xoffsets_hi,  xsquare, br ,8 , zoffsets, zoffsets_hi,4);
    Cacc1 = mac16(Cacc1, ar1 ,0, xoffsets, xoffsets_hi,  xsquare, br ,9 , zoffsets, zoffsets_hi,4);
    Cacc2 = mac16(Cacc2, ar1 ,0, xoffsets, xoffsets_hi,  xsquare, br ,10, zoffsets, zoffsets_hi,4);
    Cacc3 = mac16(Cacc3, ar1 ,0, xoffsets, xoffsets_hi,  xsquare, br ,11, zoffsets, zoffsets_hi,4);
  }

  // Read Cr from global memory
  v16int16 C0 = window_readincr_v16(DDR_IN);
  v16int16 C1 = window_readincr_v16(DDR_IN);
  v16int16 C2 = window_readincr_v16(DDR_IN);
  v16int16 C3 = window_readincr_v16(DDR_IN);

  // Convert result, add it to Cr and write back to memory
  C0=operator+(srs(Cacc0,0),C0);window_writeincr(out,C0);
  C1=operator+(srs(Cacc1,0),C1);window_writeincr(out,C1);
  C2=operator+(srs(Cacc2,0),C2);window_writeincr(out,C2);
  C3=operator+(srs(Cacc3,0),C3);window_writeincr(out,C3);
}
\end{lstlisting} 
\end{minipage}
\end{flushright}
\caption{ \textcolor{black}{Simplified version of the micro-kernel for the AIE tile.} }
\label{alg:microK_AIE}
\end{figure}

\Cref{alg:microK_AIE}
displays our micro-kernel for the VC1902. 
After some initial declarations, 
the code comprises a loop (Line~19), corresponding
to \textsf{L6}, that iterates over the
\texttt{kc} dimension of the micro-panels $A_r,B_r$.
At each iteration, the loop body
multiplies the entries in \textcolor{black}{one column} of $A_r$
(16 elements, in  \texttt{ar0} or \texttt{ar1} ) 
with those in \textcolor{black}{one  row} of $B_r$
(4 elements in  \texttt{br}, accessed through the macro
\texttt{Bref}), 
accumulating the intermediate results on four different
registers (\texttt{mac} operations in Lines 27--49).

The \ac{AIE} intrinsic \texttt{mac16()} computes 32 INT16 \ac{MAC} operations in one cycle. For the INT16 datatype, 
each \ac{MAC} operation can involve a vector with (up to) 32 elements and a second vector with (up to) 2 elements. 
However, our micro-kernel ($m_r \times n_r = 16 \times 4$) 
does not match the input dimensions of the intrinsic. 
To accommodate the dimension $n_r=4$ to the 
requirements of \texttt{mac16()},
we therefore divide the $16 \times 4$ \ac{MAC} operation into two parts;
see Lines~33 and~42.
In this manner, the additional call to the \texttt{mac16()} intrinsic increases the input by additional four times to fully utilize the function.
%
Each iteration of the loop body retrieves \textcolor{black}{four rows} of $A_r$
from the FPGA memory \textcolor{black}{(64 elements, in Lines 27--30)} and
\textcolor{black}{4 columns} of $B_r$ 
from the local memory (16 elements, in Line 24).
Therefore, a loop iteration retrieves \textcolor{black}{64+16} elements from memory levels
that are close to the arithmetic units, performing \textcolor{black}{$64 \cdot 16 \cdot 2 = 2,048$}
arithmetic operations with those. This helps to amortize the cost of memory
transfers with enough arithmetic computation, in principle yielding a 
compute-bound micro-kernel.

\textcolor{black}{Splitting the computation into two parts delivers a high computation-to-communication ratio. 
Furthermore, the high utilization of the accumulator and vector registers (respectively,
100\% and 75\% of the total resources) along 
with compiler optimization arguments facilitate to overlap the MAC operation with data transfer, thus improving performance.}

After the loop is complete, the code ``transfers'' these intermediate results
to matrix $C$.
For this purpose, 
each execution of the micro-kernel loads a $16 \times 4$ micro-tile
$C_r$ (Lines 53--56) from global memory and, after updating its contents,
stores the results back to global memory (Lines 59--63). The cost of these
final data transfers can be amortized provided $k_c$ is sufficiently large
(to be discussed later).

%% file: sections/results.tex
\section{Experimental Results}
\label{sec:results}

The evaluation of the \gemm algorithm in this section was carried out 
using the Xilinx Vitis 2022.1 developing tool.
The AIE transaction-level System C simulator was used 
to profile the timing, resource requirements, 
and assembly instructions of the designs, enabling an accurate performance analysis~\cite{2022AI_xilinx_ug1079_aieProgGuide}. 

\subsection{Approximate computing with different datatypes}
\label{subsec:datatype}
Precision scaling is a prominent approximate computing technique 
that is often leveraged in quantized DL inference on edge devices in order to
reduce energy consumption while maintaining acceptable precision in the results.
In this line, the Versal VC1902 supports arithmetic for multiple datatypes, 
which can be exploited to improve the cost-effectiveness ratio.
\textcolor{black}{Concretely, the Versal provides a variety of real datatype precision formats.
Concretely, the AIE supports INT8, INT16 and FLOAT32, yet 
the performance of the different formats varies significantly, with lower precision requiring fewer AIE cycles.
For example, in one clock cycle, the ac{AIE} can
perform 128 UINT8 MACs operations; 32 for INT16, and 8 for FP32.
}

These figures motivate us to choose 
\textcolor{black}{INT16} as the baseline datatype for our DL-oriented \gemm kernel on the Versal
VCK190, as this
\textcolor{black}{datatype provides fair precision, reducing memory utilization and accelerating computation.
In addition, the \texttt{mac16()} intrinsic for the INT16 datatype accommodates a flexible implementation.
The general design of the micro-kernel will be adapted  for lower precision in future work.}

 \subsection{Evaluation of the micro-kernel for the AIE tile}
\label{subsec:micro_kernel_design}

In subsection~\ref{subsec:micro_kernel_design} we motivated the selection
of an $m_r \times n_r = 16 \times 4$ micro-kernel (for the \textcolor{black}{INT16} datatype).
While there exist other micro-kernel dimensions that will also
produce correct results, 
choosing a smaller micro-kernel results in 1) using fewer accumulator
registers/lanes which will likely render idle 
cycles for the accumulator arithmetic units yielding lower performance; and 2) 
a worse arithmetic-memory ratio for the body of loop \textsf{L6}. Furthermore,
the compiler can transparently deal with 
a micro-kernel that exceeds the maximum number of 
registers/lanes by 
temporarily saving 
certain data to memory (and eventually restore it) 
during the micro-kernel execution. %
Unfortunately, this register spilling
can significantly 
degrade performance when occurring inside loop \textsf{L6}.

To illustrate this, we performed the following two experiments.
In the first case, we profiled the execution of our
$16 \times 4$ micro-kernel with \textcolor{black}{$k_c$ = 256}. 
The number of  \textcolor{black}{MAC} operations performed
inside the micro-kernel for this configuration is thus
\textcolor{black}{16,384, and the design took 596 cycles to complete, which corresponds to 27.5 MACs per cycle.}
The theoretical peak for the AIE tile when operating with \textcolor{black}{INT16 and accumulating
on the 16 lanes per accumulator register is 32 MACs per cycle.
Here, the small performance drop due to the cost of loading
the micro-tile $C_r$ from the global memory, updating it, and storing
the results back in memory.}

In the second configuration, we employed a $32 \times 4$ micro-kernel which,
in theory, features a higher re-use rate of $C_r$ while this
micro-kernel resides in the local
memory.
However, this approach exceeds the capacity of the 
accumulator registers, producing register spilling.
Concretely, \textcolor{black}{with $k_c$ = 256}, this design performs twice as many MAC operations
as the previous one, and should thus offer a theoretical cost 
that roughly corresponds to
 \textcolor{black}{$596\cdot 2 = 1,192$ cycles.
In practice though, it required
1,429 cycles to complete, that is, about 20\% longer than expected.
The second configuration thus only achieves 23 MACs per cycle, which is far from the AIE peak.}

We close the previous discussion with two final remarks:
\begin{itemize}
\item \textcolor{black}{%
    The accumulator registers in the  Versal VC1902 can be viewed as  \textcolor{black}{four 768-bit accumulators.} Our micro-kernel involves 8 \texttt{mac16()} intrinsics that operate with these 4 independent accumulator registers, therefore
    using 100\% of this type of resource. Maintaining the INT16 data type while augmenting the size of micro-kernel could
    increase the utilization of the register space, but this would eventually lead to register spilling.
    The current design utilizes 75\% of the space from the vector register resources. Indeed, the second scenario presented in this section (with a $32 \times 4$ micro-kernel) exceeds the maximum vector register resources. In consequence, during the execution of the micro-kernel loop, some of these extra vector need to be temporally 
    stored in the local memory to be later retrieved back into vector registers, leading to a performance degradation that
    we aim to avoid.~\cite{UG1079}
      }
\item The idea behind the 48-bit accumulators is to have 16-bit multiplication
      results and accumulate over those results without bit overflows. 
      In the VCK190, 48-bit is the minimum accumulator ``size''.
\end{itemize}

\subsection{Selecting the cache configuration parameter $k_c$}
\label{subsec:impact_of_KC}

Having determined the dimensions of the micro-kernel, we
next investigate how to select the optimal value for $k_c$.
Note that we proceed through the hierarchy from the processor registers
toward the slower levels of the memory system. Thus, in the previous 
subsection we discussed the values of
$m_r\times n_r$ in connection with the number of accumulator registers; 
in this subsection we choose $k_c$ taking into account the capacity
of the local memory; and in the next subsection we will investigate
how to set the optimal value of $m_c$ 
with respect to the capacity of the FPGA RAMs.

Ideally, we would like $k_c$ to be as large as possible because,
from the point of view of the micro-kernel, 
that option reduces the overhead of
loading/saving the micro-tile $C_r$ from/to global memory
($m_r \times n_r$ reads plus the same number of writes from that level)
compared with the number of arithmetic operations
($2 \, m_r \cdot n_r \cdot k_c$).
Also, the micro-panel $B_r$ that resides in the local memory 
is $k_c \times n_r$ and a larger $k_c$ implies a better re-use
of the resources in that level. However, with $n_r=4$ already set,
the largest dimension for $k_c$ is limited by the capacity of the
local memory. 
Furthermore, as $k_c$ is the blocking parameter
for the $k$-dimension of \gemm, from the application perspective
we have the constraint that $k_c \le k$.

\begin{figure}[htbp]
    \centering
    \includegraphics[width=0.5\columnwidth]{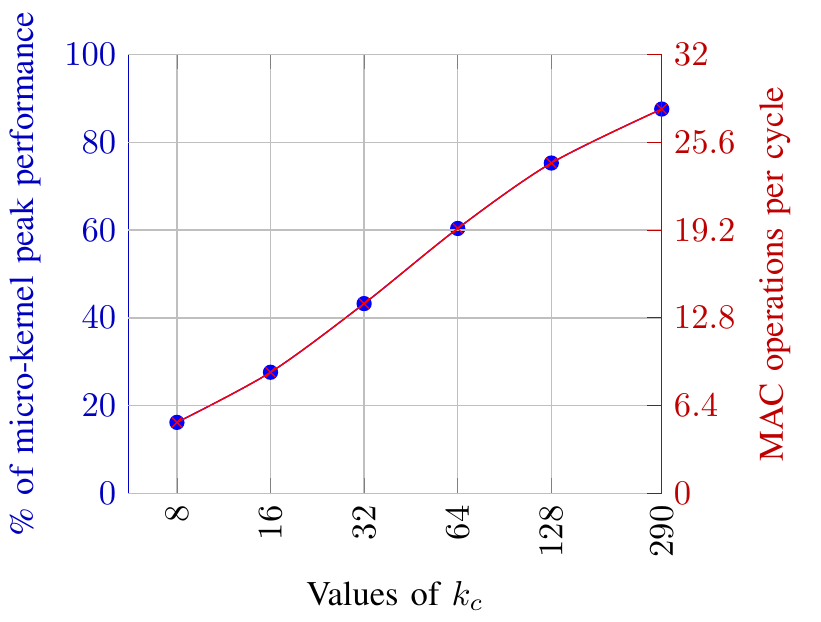}
    \vspace*{-2ex}
    \caption{\color{black}{Impact of $k_c$ on the performance of the $m_r \times n_r = 
	16 \times 4$ micro-kernel. } }
    \label{fig:kc_sweep}
	
\end{figure}

In this subsection, we first demonstrate the impact
of $k_c$ on the performance of the micro-kernel.
For that purpose, 
we profiled
the ratio of cycles spent in MAC operations (useful
arithmetic) against the total 
number of cycles (including the overhead due to loading/storing
data) when the micro-kernel is executed in isolation. 
\Cref{fig:kc_sweep} illustrates the effect of this parameter on
performance.
For instance, with \textcolor{black}{$k_c=64$}, the micro-kernel 
required a total of \textcolor{black}{212} cycles to execute, 
out of which \textcolor{black}{150} correspond
to MAC operations and the remaining are due to
data transfers. 
Thus the efficiency of this configuration is around \textcolor{black}{60\%}.
As $k_c$ is increased, it approaches an 
horizontal asymptote:
\textcolor{black}{ 
75.3\%, 
85.9\%,
and 87.6\%
for $k_c=$ 128, 256 and 290,} respectively. 
When $n_r=4$, at most $k_c=290$ %
in order to ensure that the micro-panel $B_r$ fits into the local memory.

This experiment shows that the ratio between 
arithmetic 
and data transfers for the micro-kernel strongly depends on the value of the cache configuration
parameter $k_c$.
In particular, the copies 
for $C_r$ from the global memory  to local memory 
can be basically hidden by choosing an appropriate (large)
value for $k_c$.
In addition, choosing a proper dimension of the micro-kernel helps
to reduce the negative impact of the copies 
for $B_r$ from the local memory and the micro-panels of $A_c$ from the FPGA
memory to about  \textcolor{black}{ 12\% } of the cycles. 

Here it is important to emphasize that, once $k_c$ is fixed, 
this sets an upper bound on the performance of the \gemm kernel,
since the complete algorithm is assembled using the micro-kernel as
the cornerstone building block.
The only missing data transfers that we did not consider yet, because
they do not occur inside the micro-kernel, correspond to the copies from the buffer
$B_c$, in global memory, to the micro-panel $B_r$, in the local memory; 
see \Cref{fig:systemDiag}.
These copies are orchestrated by the scalar engines (ARM processors) in the VCK190
and their cost is discussed in the following subsection.

\ 

\subsection{Selecting the cache configuration parameter $m_c$}
\label{subsec:impact_of_MC}

Consider now the buffer $A_c$ that resides in the FPGA
RAMs, of dimension $m_c \times k_c$,
and let us follow a reasoning analogous to that applied for $k_c$.
First, there are two upper bounds on $m_c$: 1) With $k_c$ set in the previous subsection,
the largest dimension for $m_c$ can be easily derived from the capacity of the
FPGA memory; and 2) from the application point of view,  
as $m_c$ is the blocking parameter from the problem dimension $m$, we have that
$m_c \le m$.
Let us discuss next the benefits of choosing a large value for $m_c$ 
by analyzing the overhead due to a data copy from the buffer $B_c$ in
the global memory to the micro-panel $B_r$ in the AIE tile local memory.
For this purpose, consider for example that $k_c=$ \textcolor{black}{290} and $n_r=$ 4.
For this configuration, 
the data copy between global memory and the local storage introduces a significant
overhead: \textcolor{black}{8,309} cycles for the copy versus 
\textcolor{black}{8,952  cycles for the execution of a single micro-kernel. 
The number of MAC operations per cycle is thus reduced to only 2.}
Fortunately, this type of data copy only occurs 
once per iteration of loop \textsf{L4}, and the copied micro-panel $B_r$ is then re-used
in $m_c/m_r$ executions 
of the micro-kernel
(one per iteration of loop \textsf{L5}).
Therefore, with $m_r$ already fixed, 
this reasoning points in the direction of choosing a large
value for $m_c$, 
because a large ratio $m_c/m_r$ then paves the road toward amortizing
the cost of the copy for $B_r$ with enough arithmetic from inside the micro-kernel.

\begin{figure}[htbp]
    \centering
    \includegraphics[width=0.5\columnwidth]{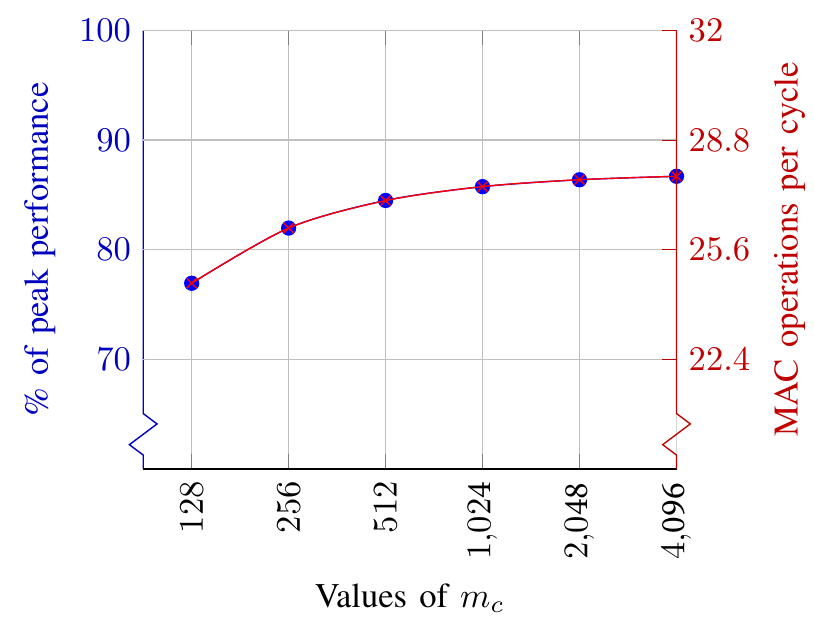}
    \vspace*{-2ex}
    \caption{ \color{black}{Impact of $m_c$ on the performance of the \gemm kernel. For this 
     experiment, 
     $m_r \times n_r=16 \times 4$,
     $(m,n,k)=$ (4,096, 4,096, 290), 
     $n_c=n$, and $k_c=k$.}
    }
    \label{fig:mc_sweep}
\end{figure}

To close this section, we explore the experimental impact of $m_c$ on the performance
for a specific \gemm problem. For that purpose, 
we chose $(m,n,k)=$ (4,096, 4,096,  \textcolor{black}{  290}), set
$n_c=n,k_c=k$, and varied $m_c$ for different executions of the \gemm kernel.
 \Cref{fig:mc_sweep} shows the performance benefits from choosing
a larger value for $m_c$, but at the same time exposes 
a clear horizontal asymptote around \textcolor{black}{87.6\%} 
 of the peak performance (which corresponds to \textcolor{black}{32 INT16 MACs, or 64 INT16 arithmetic operations,}
per cycle). 
Therefore, 
this is perfectly aligned with the performance of the micro-kernel when considered
in isolation (that is, the building block for our \gemm kernel), 
which in \Cref{fig:kc_sweep} was reported to achieve up to \textcolor{black}{86.7\%}of the
peak performance.
This implies that the overhead due to the copies between $B_c$ and $B_r$
can be hidden with arithmetic provided $m_c$ is chosen to be large enough.

 As a rough summary of the previous analysis, our realization of \gemm for the Versal
losses \textcolor{black}{15\%} of the peak performance because of the copies from
$A_c$ and $B_r$ to the vector registers, while an additional 
4--5\% is lost because of the transfers between $B_c$ and $B_r$. 